\documentclass[a4paper,fleqn,usenatbib,useAMS,referee]{mnras}
\usepackage{txfonts}
\usepackage{epsfig}
\usepackage{hyperref}
\usepackage{xspace}
\usepackage[T1]{fontenc}
\usepackage{ae,aecompl}
\usepackage{multirow}
\usepackage{booktabs}
\usepackage{times}
\usepackage{caption}
\usepackage{subcaption}
\usepackage{{threeparttable}}
\usepackage{multicol}

\usepackage{graphicx}	% Including figure files
\usepackage{amssymb}	% Extra maths symbols
\usepackage{lscape}
\usepackage{placeins}
\usepackage{soul}
\usepackage{longtable}
\usepackage{array, booktabs, longtable, makecell}
\usepackage{rotating}
\usepackage{pdflscape}	% Landscape pages

\newcommand{\tes}{\textit{TESS}}

\newcommand{\jktabs}{\textsc{jktabsdim}}

\newcommand{\ravespan}{\textsc{ravespan}}

\newcommand{\isp}{\textsc{ispec}}

\newcommand{\sigspec}{\textsc{sigspec}}

\newcommand{\kms}{\,km\,s$^{-1}$}
\newcommand{\U}{\textit{U}}
\newcommand{\B}{\textit{B}}
\newcommand{\V}{\textit{V}}
\newcommand{\R}{\textit{R}}
\newcommand{\I}{\textit{I}}
\newcommand{\J}{\textit{J}}

\newcommand{\K}{\textit{K}}

\newcommand{\bstargrid}{\textsc{TLUSTY\ BSTAR2006}}
\newcommand{\ostargrid}{\textsc{TLUSTY\ OSTAR2002}}

\newcommand{\gaia}{\textit{Gaia}}

\newcommand{\beq}{\begin{equation}}
\newcommand{\eeq}{\end{equation}}
\title[$\beta$ Cephei pulsators in eclipsing binaries]{Selected $\beta$ Cephei pulsators in eclipsing binaries}

\author[ Ö. Çakırlı et al. ]{Ö. Çakırlı$^{1,3}$\thanks{E-mail: \href{mailto:omur.cakirli@ege.edu.tr} {omur.cakirli@ege.edu.tr}}, 
B. Hoyman$^{1,2}$, O. Özdarcan$^{1}$ and S. Bilir$^3$\\
$^1$Ege University, Science Faculty, Astronomy and Space Science Dept., 35100 Bornova, {\.I}zmir, Türkiye \\
$^2$Atatürk Üniversitesi, Fen Fakültesi, Astronomi ve Uzay Bilimler Bölümü, 25240 Yakutiye/Erzurum, Türkiye \\
$^3$Istanbul University, Faculty of Science, Department of Astronomy and Space Sciences, 34119, Beyazıt, Istanbul, Türkiye
}
\date{Accepted XXX. Received YYY; in original form ZZZ}

\pubyear{2025}

\begin{document}
\label{firstpage}
\pagerange{\pageref{firstpage}--\pageref{lastpage}}
\maketitle

% Abstract of the paper
\begin{abstract}
We present a comprehensive analysis of five eclipsing binary systems containing $\beta$\,Cephei-type 
pulsating components. These systems were identified using the high-precision Transiting Exoplanet Survey Satellite 
(\tes) photometry and complemented by high-resolution spectroscopic data from multiple instruments (e.g., HARPS, 
FEROS, UVES). For each target, we derived fundamental parameters including masses, radii, effective temperatures, 
and metallicities through a combined analysis of light curves and radial velocity data. The atmospheric parameters 
were determined via spectral disentangling and MCMC-based fitting techniques. Absolute parameters were used to 
position components on the HR diagram, and evolutionary status was assessed using MESA/MIST stellar 
tracks. Frequency analysis revealed multiple significant pulsation modes in all systems. Our study increases 
the number of well-characterized $\beta$\,Cephei pulsators in binaries and provides high-precision benchmarks 
for stellar evolution and asteroseismology.
\end{abstract}
% Don't make up new ones.
%%%%%%%%%%%%%%%%%%%%%%%%%%%%%%%%%%%%%%%%%%%%%%%%%%

%%%%%%%%%%%%%%%%% BODY OF PAPER %%%%%%%%%%%%%%%%%%
\section{Introduction} \label{sec:intro}
\label{intro}
Understanding massive star evolution is essential for an extensive range of topics within stellar astrophysics. The field of 
asteroseismology is one of the most significant of these topics. Asteroseismology is crucial for understanding the interior 
physics and structure of stars through their pulsations \citep{2021RvMP...93a5001A}. It also presents a unique method for 
examining the properties of massive stars, which exhibit pulsations \citep{2019ApJ...883L..26B,2020FrASS...7...70B}, and is 
particularly compelling due to its significant potential for future research.

There are three known types of massive star pulsators: $\beta$\,Cephei \citep[$\beta$\,Cep;][]{1993SSRv...62...95S,
2003SSRv..105..453A,2005ApJS..158..193S, 2013pss4.book..207H}, slowly pulsating B type stars \citep[SPB;][]{1998A&A...330..215W}, and 
a third group characterised by stochastic low-frequency variability \citep[SLF;][]{2020A&A...640A..36B}. As of today, we have not yet fully determined the boundaries for all types on the Hertzsprung-Russell (HR) diagram, which is evident in some studies 
\citep{2023Ap&SS.368..107B,2022A&A...659A.142S}. Massive stars are mainly observed in multiple systems \citep[e.g.,][]{2012Sci...337..444S,2022MNRAS.513.3191S}; therefore, a significant portion is expected to exist in eclipsing binaries usually as primary components. By modelling eclipsing binaries, which are described as the royal road to stellar astrophysics \citep[][]{1948HarMo...7..181R,2005Ap&SS.296....3B,2012ocpd.conf...51S}, model-independent, precise stellar parameters (e.g., radius, mass), which serve as essential calibrators for stellar evolution theory \citep[][]{2010A&ARv..18...67T,2019ApJ...872L...9P}, are obtained. Dynamical masses derived from eclipsing binary modelling, high-resolution spectroscopy, and asteroseismic investigations provide significant chances to study the physics of stellar evolution models further and address the complex and peculiar discrepancies between observations and models \citep{2020A&A...637A..60T}. A significant advantage is that the combined study of photometry and spectroscopy may assist us in precisely determining the positions of the components of the binary system on the HR diagram \citep{2024PARep...2...41E,2025PARep...3...43E}. This approach may allow us to clearly define the boundaries of massive pulsators in the HR diagram. However, the total count of reported massive stars in eclipsing binary or multiple systems is considerably lower than that of their low-mass counterparts \citep{2016AJ....151...68K}. Various observations from space attempt to change the current report regarding the topic. The significant limitation of the small sample size has, to this point, restricted wider asteroseismic knowledge of massive stars. Researchers have been focused since the beginning of the Transiting Exoplanet Survey Satellite mission \citep[\tes;][]{2015JATIS...1a4003R} to collect a comprehensive and adequate number of samples of massive pulsators for the study of massive star asteroseismology by analysing the \tes\ data, one of the most significant space-based surveys \citep[e.g.,][]{2019ApJ...873L...4H}.

As a result, several authors have recently released catalogues of massive stars based on these studies 
\citep[][] {2020A&A...639A..81B,2019ApJ...872L...9P,2024ApJS..272...25E}. These studies aim to provide a 
reliable sample for future asteroseismology by integrating \tes\ photometry with spectroscopy and advancing 
asteroseismic modelling of single or binary $\beta$\,Cephei stars. Furthermore, due to the advantages 
of eclipsing binary modelling and asteroseismology, numerous recent investigations have either focused on 
massive pulsators within eclipsing binary systems or expanded their focus to include a larger number of more 
massive eclipsing binaries in the sample 
\citep[][]{2015EPJWC.10104001S,2020MNRAS.497L..19S,2021MNRAS.501L..65S,2021A&A...652A.120I,2021A&A...650A.112Z,2022MNRAS.513.3191S}. 

While the primary aim of some of these authors was to identify eclipsing binaries, their contributions remain valuable for 
identifying $\beta$\,Cephei pulsators within these systems, as they provide a compiled list of existing eclipsing binaries. It 
is important to highlight that nearly all of these studies rely on light curve analysis, representing the first comprehensive 
photometric examination of $\beta$\,Cephei-type stars in massive binaries and clearing the way for a significant investigation 
of their positioning on the HR diagram.

The selected systems have been subjected to data collection, including time-series \tes\ data and high-resolution spectroscopic 
data (see to \S~\ref{data}). These data will be utilised to conduct thorough atmospheric and photometric modelling 
(\S~\ref{analysis} and \ref{lc}) and accurately determine the stellar parameters. In \S~\ref{sec_iso}, we present the evolutionary 
models that we employed. \S~\ref{pulsi} provides a comprehensive analysis of the pulsation characteristics of our samples. The 
study concludes with a discussion of the key findings, presented in \S~\ref{conc}.
\begin{table*}
\centering
\caption{Published information regarding the targets investigated in this work.}
\begin{tabular}{lccccc}
	\hline
	\hline
Parameters 						 &TIC\,426520557&TIC\,93997745&TIC\,339570153 &TIC\,264540478& TIC\,379012185\\
\hline
Common Name     				 &HD\,254346	&HD\,74455    &CPD\,-41\,7746 &$\psi$\,Ori    &EK\,Cru            \\
RA (J2000) (h:m:s)    			 &06:16:57.32 	&08:42:16.19  &16:54:29.49    &05:26:50.22 	  &12:02:58.47      \\
DEC (J2000) ($^{\circ}$:$'$:$''$)&+22:11:41.96  &-48:05:56.75 &$-$41:39:14.95 &+03:05:44.42   &$-$62:40:19.23   \\
$V$ (mag)           			 &9.74$^a$		&5.47$^a$     &9.19$^b$       &4.61$^c$		  &8.11$^d$             \\
\hline
Sectors                 		 &43-45,71-72	&8-9,35-36,61-63	&39,66    &6,32           &10,11,37,38,64 \\
\multirow{2}{*}{Observation Start/End (UT)}&16/09/2021 &01/03/2019 &05/27/2021&15/12/2018&03/26/2019\\ 	
										   &07/12/2023 &12/02/2023 &06/26/2023&16/12/2020&05/03/2023\\
exptime/type 							   &120/SPOC   &120/SPOC   &120/SPOC  &120/SPOC  &120/SPOC  \\
\hline
N$_{u}$      							   &--- &2   & --- &6  & ---\\
N$_{h}$      							   &--- &12  & --- &---& ---\\
N$_{f}$      							   &--- &--- & --- &---& 5  \\
N$_{g}$      							   &--- &--- & 76  &---& ---\\
N$_{s}$      							   &12  &--- & --- &1  & ---\\
N$_{e}$      							   &--- &--- & --- &29 & ---\\
%N$_{x}$      							   &--- &--- & --- &---& ---\\
\hline	
\end{tabular}\\
\footnotesize
The information is based on the data accessible on MAST. The TIC numbers are utilised to organise 
the systems in this list. The \tes\ observing baseline includes the start and end dates, listed in the "Observing 
start/end" line. It is worth mentioning that the systems may not have been observed continuously throughout this 
entire period. The light curves of all systems are obtained using a 120-second exposure time cadence, as noted 
by the Science Processing Operations Centre \citep[SPOC;][]{2016SPIE.9913E..3EJ}. The final part indicates the 
quantity of optical spectra utilised for each spectrograph: 
%
%\red{N$_x$: XSHOOTER}; 
%
N$_e$: ELODIE; N$_s$: SOPHIE; N$_h$: 
HARPS; N$_u$: UVES; N$_f$: FEROS and N$_g$: GIRAFFE. References of the $V$ (mag): $^a$: \citet{2000A&A...355L..27H}, 
$^b$: \citet{2002A&A...384..180F}, $^c$: \citet{2013AJ....145...37S}, 
$^d$:\citet{2001A&A...374..204D}, $^e$:\citet{1997ESASP1200.....E}.
\label{systems}
\end{table*}

\section{Observations}\label{data}
\subsection{Photometry}\label{photometry}
Between July 2018 and October 2024, light curves from the 82 sectors of the \tes\ were collected and made 
accessible for the many objects. Reduced light curves are obtained from the data by the Science Processing 
Operations Center \citep[SPOC;][]{2016SPIE.9913E..3EJ} and made available through the Mikulski Archive 
for Space Telescopes (MAST) portal\footnote{\url{https://mast.stsci.edu/portal/Mashup/Clients Mast/Portal.html}}. We 
used the simple aperture photometry (SAP) light curves in this work. We visually inspected the \tes\ light 
curves of approximately 2\,500 eclipsing binaries present in a bibliography maintained by the authors and identiﬁed 
five objects that show both eclipses and previously unrecognized $\beta$\,Cephei pulsations. CPD\,-41\,7746, $\psi$\,Ori 
and EK\,Cru had been previously identified as $\beta$\,Cephei pulsators \citep[see e.g.,][]{2024ApJS..272...25E} but 
their absolute parameters and detailed pulsational characteristics have not been determined.

Light curve data with a 2-minute cadence is typically supplied by simple aperture photometry (SAP) 
and pre-search data conditioning simple aperture photometry (PDCSAP). The analysis focuses on SAP flux, with 
long-term trends eliminated through the application of Co-trending Basis Vectors (CBVs). PDCSAP flux typically 
provides cleaner data compared to the SAP flux and exhibits fewer systematic trends. For instance, in their 
research done within \citet{2024MNRAS.531.3823Y} and \citet{2023NewA..10102022Y}, they analysed and compared 
SAP and PDCSAP data for particular eclipsing binaries, namely V757\,Cen and PP\,Lac. Some PDCSAP light 
curves of the systems have been deleted, leading to the loss of times of minima. Upon considering all examples 
and examining variations in the light curves of our systems through each sector, we adopted SAP fluxes, 
despite the presence of instrument effects. In the analysis of light curves, we use the complete set of 
SAP data for the systems.

\begin{figure*}
	\center
\includegraphics[width=1\textwidth]{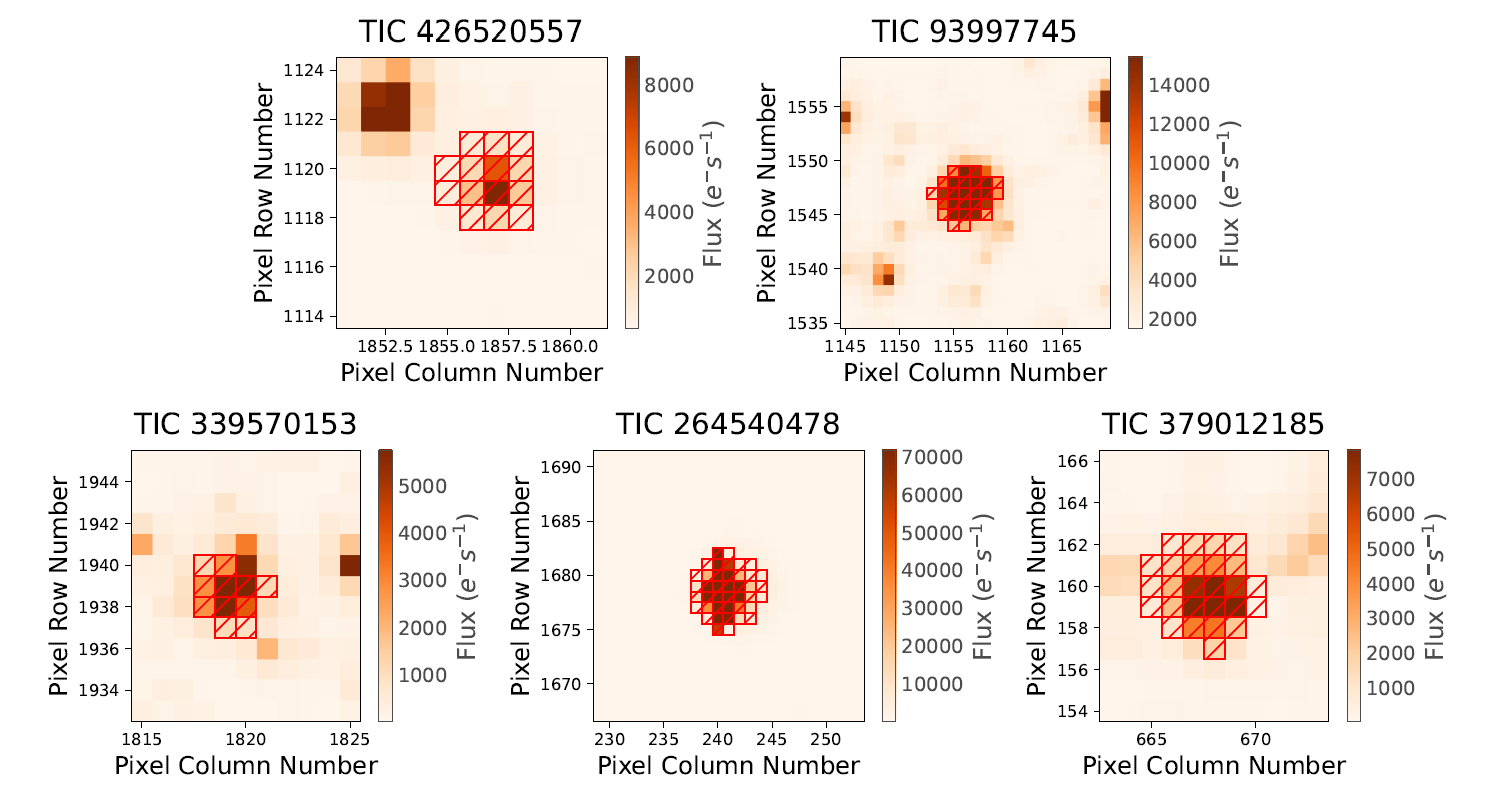}   
\caption{A graphical representation of the \tes\ target pixel files for the systems mentioned in this 
\S\ref{photometry}. The stars in challenge were those for which custom light curves were created. The panel shows the target pixel ﬁles, with the star of interest shown in the centre and the potentially contaminating stars around it with the pipeline aperture marked with a cross-hatched area.
}
	\label{TPFs}
\end{figure*}

We used the \textsc{lightkurve} \citep{2018ascl.soft12013L} to plot the Target Pixel Files 
(TPF; in Figure\,\ref{TPFs}) for each target star in the sample, aiming to investigate potential blends and 
sources of contamination in the apertures. The analysis of the TPFs is especially important in densely 
populated areas, given that the relatively large pixel size of the \tes\ CCDs (21 arcseconds per pixel) 
can lead to significant photometric contamination from adjacent sources. Following the target selection, we 
carefully reassessed the quality of the reduced SAP light curves. First, we inspected the aperture masks 
used in the pixel data and evaluated both the level of captured ﬂux and the level of contamination from 
nearby stars. 

Table\,\ref{systems} presents the detailed observational data for five binaries, including the sector 
number, the observation time, the exposure time, and the type of light curve. We use 25-sector observations 
from \tes\ to determine eclipse times and analyse the small intrinsic variability of light curves. SAP fluxes 
were normalised for each sector and converted to normalised flux for purposes of analysis.
%

%
%
%%%%%%%%%%%%%%%%%%%%%%%%%%%%%%%%%%%%%%%%%%%%%%%%%%%%%%%%%
%Figure 01
%%%%%%%%%%%%%%%%%%%%%%%%%%%%%%%%%%%%%%%%%%%%%%%%%%%%%%%%%%%
%
%

\subsection{Spectroscopy}\label{spectroscopy}
This section outlines the spectroscopic observations of our targets utilising the Ultraviolet and Visual Echelle 
Spectrograph \citep[UVES;][]{2000SPIE.4008..534D}, the High Accuracy Radial-velocity Planet Searcher 
\citep[HARPS;][]{2003Msngr.114...20M}, the Fiberfed Extended Range Optical Spectrograph \citep[FEROS;][]{1999Msngr..95....8K}, 
the high-resolution Fibre Large Array Multi Element Spectrograph \citep[FLAMES/GIRAFFE;][]{2002Msngr.110....1P}, the high 
resolution/high-precision/fiber-fed spectrograph \citep[SOPHIE\footnote{\url{http://atlas.obs-hp.fr/sophie/}};][]{2008SPIE.7014E..0JP}, 
and the fiber-fed echelle spectrograph \citep[ELODIE\footnote{\url{http://atlas.obs-hp.fr/elodie/}};][]{1996A&AS..119..373B}. The 
spectral resolution of the quoted spectrographs for the instrument settings used is: 
%
%R$\sim$15\,000 (XSHOOTER), 
%
$\R\sim$48\,000 (FEROS), $\R \sim$115\,000 (HARPS), $\R \sim$40\,000 (UVES), $\R \sim$75\,000 (SOPHIE), $\R \sim$42\,000 (ELODIE), 
and $\R \sim$6\,300 (GIRAFFE). 
%\red{[XSHOOTER ile ilgili kısma gerek kalmadı, kaldırılabilir.]}
%

These instruments were employed to conduct atmospheric analysis of the component stars and extract the radial velocity (RV) 
measurements. For further information about the wavelength ranges of the spectra or the resolution mode, we guide you to the 
ESO web pages (Science Archive Facility\footnote{\url{http://archive.eso.org/wdb wdb/adp/phase3_main/form}}). The information 
regarding the total number of spectra can be found in Table\,\ref{systems}. The quoted spectra were obtained by using the 
available spectra that were collected by the program IDs; 099.C-0637(A)(PI: Nieva Maria-Fernanda), 072.D-0286(A)(PI: A.P. Hatzes), 082.B-0610(A)(PI: J. Simmerer), 087.B-0308(A)(PI: P. Coelho), 089.D-0975(A)(PI: R. Barba), and 099.D-0895(A)(PI: H. Sana).

\section{Analysis methods}\label{analysis}
\label{sec:anal}
\subsection{Radial velocity measurements and orbital solutions}
\label{ravespan}
In our spectroscopic data survey utilising various instruments, we assess the RVs of binary system components 
through the \ravespan\ code \citep{2013MNRAS.436..953P,2015ApJ...806...29P}. All extracted RVs described in the paper 
are derived from the radial velocity profiles obtained through the analysis of individual spectra using the linear 
Broadening-Functions deconvolution technique \citep{2002AJ....124.1746R}. The technique requires greater computational 
power compared to the more straightforward and inherently non-linear Cross-Correlation Function (CCF) technique. It 
offers profiles concerning the shape and strength of the spectral lines for a star of the same spectral type. The 
templates were interpolated using grids of theoretical stellar spectra. Advancements in the analysis of certain 
systems led to enhancements in surface temperatures and gravities, which were subsequently employed to recalculate 
the templates and to renew the radial velocities. With the measured radial velocities, we made a preliminary model 
of the target systems and obtained the binary component parameters. Then, we generate a new synthetic template spectrum 
with the improved binary component parameters and measure radial velocities with the improved template. Applying this 
procedure iteratively a few times, we obtained final radial velocities tabulated in Table~\ref{tab:target_RVs}.

\begin{figure*}
\includegraphics[width=1\textwidth]{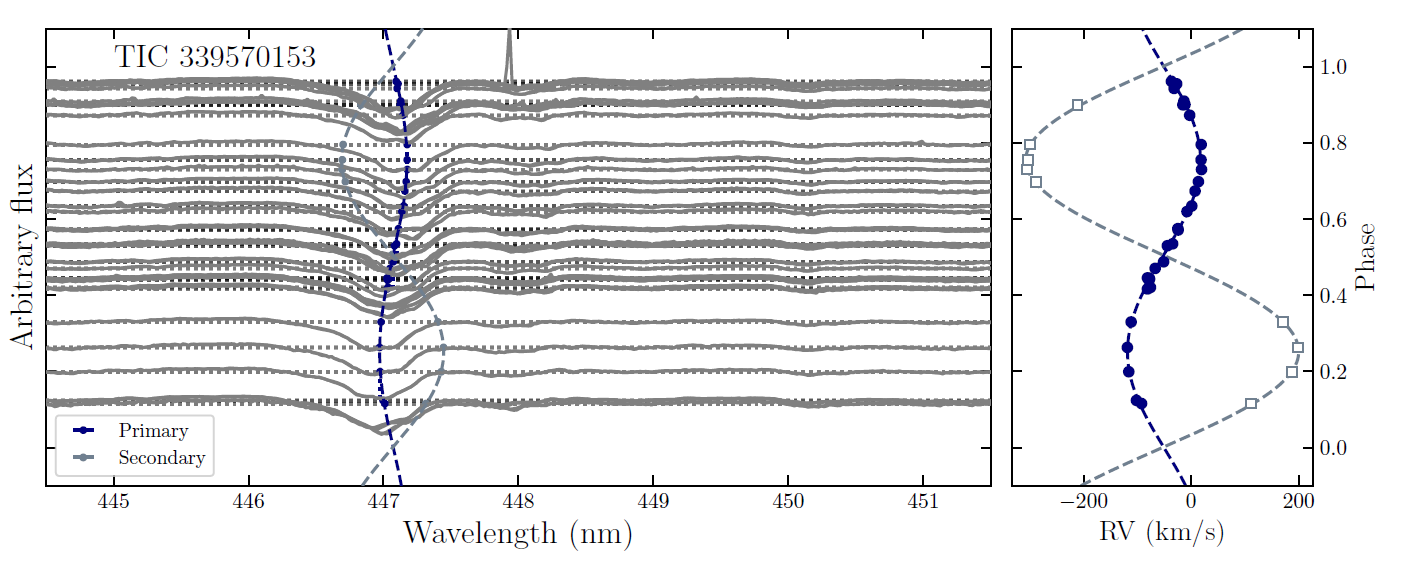}
\caption{The left panel shows a set of absorption lines involving a secondary component, which matches the orbital period cycle of 
the system. The variation of the lines is represented by dashed lines. The right panel displays the radial velocity of the absorption 
lines. The filled circles on the radial velocity plot indicate the primary, whereas the open squares represent the secondary.}
	\label{phased_mcmc_rv}
\end{figure*}

We can identify separate peaks belonging to both components in the graph of the broadening function for each system. We 
apply a \textit{Gauss} function to each peak individually and determine the peak position of the function along the 
velocity axis. The peak position is regarded as the measured radial velocity of the corresponding component. 
Figure~\ref{phased_mcmc_rv} clearly demonstrates the variation in the absorption lines for the TIC\,339570153 
system. The figure illustrates the variation in the phase of the isolated He\,{\sc i} $\lambda$ 4471 line 
profiles throughout multiple runs. The absorption lines from the system are displayed and shifted along the orbital 
phases. The panel located on the right side of Figure~\ref{phased_mcmc_rv} presents the measured radial velocities 
along with the fitted radial velocity model. We were successful in estimating the radial velocity measurements 
for each component of every system. The results are available in Table~\ref{tab:target_RVs}.
%

%
%   
%%%%%%%%%%%%%%%%%%%%%%%%%%%%%%%%%%%%%%%%%%%%%%%%%%%%%%
%Figure 02
%%%%%%%%%%%%%%%%%%%%%%%%%%%%%%%%%%%%%%%%%%%%%%%%%%%%%%
%
%

%
%
%%%%%%%%%%%%%%%%%%%%%%%%%%%%%%%%%%%%%%%%%%%%%%%%%%%%%%%%%
%\begin{figure*}
	%	\center
%\includegraphics[width=14.5 cm]{TIC_364398410_420.0-460.0nmRegion_10000xMCMC_Corner_forPaper.pdf}
%\caption{Example corner plot showing posterior probability distribution for the atmospheric parameters 
%from 48 walkers $\times$ 10\,000 step simulations for TIC\,364398410, determined using the code. Contour 
%levels correspond to 1, 2, and 3$\sigma$, and the histograms on the diagonal represent the posterior 
%distribution for each parameter, with the mode and internal 99\% confidence levels indicated. More 
%realistic errors are discussed in the text.
%	}
%	\label{mcmc_rob_e}
%\end{figure*}
%%%%%%%%%%%%%%%%%%%%%%%%%%%%%%%%%%%%%%%%%%%%%%%%%%%%%%%%%%
%
%

%
%
%%%%%%%%%%%%%%%%%%%%%%%%%%%%%%%%%%%%%%%%%%%%%%%%%%%%%%%%%
%Figure03
\begin{figure*}[h]
\center
\includegraphics[width=1\textwidth]{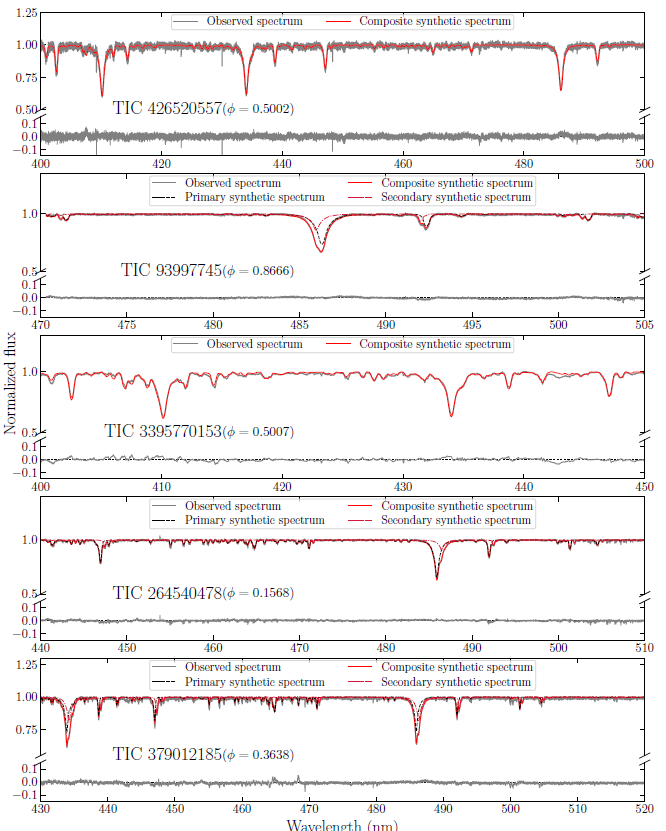}
\caption{The observed spectra of the systems are compared to composite synthetic spectra (solid lines). The primary and secondary components are represented by dashed and dash-dotted lines, respectively.
}
	\label{mcmc_tayf}
\end{figure*}
%%%%%%%%%%%%%%%%%%%%%%%%%%%%%%%%%%%%%%%%%%%%%%%%%%%%%%%%%%
%
%

%
\subsection{Atmospheric parameters}
\label{spec_analysis}
A significant challenge encountered when trying to match synthetic spectra with those observed from 
double- and single-lined spectroscopic binary systems is that they consist of a blend of the signals 
from both components. By making use of multiple well-phase-covered observational epochs with the highest 
signal-to-noise ratio (SNR), it is feasible to separate the signals from the two components and obtain 
separated spectra for each. In this study, we select our code that employs the techniques multiple times 
to improve the independent atmospheric parameters alongside nearly precise initial parameters and the 
radial velocities of the components to separate the spectra \citep{2024MNRAS.tmp.1906C}. The procedure 
we use is very similar to that employed in earlier studies 
\citep[see, e.g.][]{2023MNRAS.523.1676C,2024MNRAS.533.2058C}. In summary, our code which was developed in the 
\textsc{python} programming language, employs \isp\ subroutines \citep{2014A&A...569A.111B, 2019MNRAS.486.2075B} 
and pre-generated synthetic spectrum libraries that accompany the \isp\ code to create a synthetic spectrum for 
each component. During solutions, depending on the effective temperature, it might be necessary to transition between 
different grids. As a result, for $15\,000 \leq\,\rm T_{eff}<30\,000\,K$ solutions \bstargrid\ grid, and for 
$\rm T_{eff}\,\geq 30\,000 K$ solutions \ostargrid\ grid were used. \bstargrid\ and \ostargrid\ grids were BSTAR 
\citep{2007ApJS..169...83L} and OSTAR \citep{2003ApJS..146..417L} grids generated using the \textsc{tlusty} code 
\citep{1988CoPhC..52..103H, 1995ApJ...439..875H} and modified to be compatible with \isp\ software. Additionally, the 
code utilises Markov Chain Monte Carlo methods (MCMC; \citealt{2010CAMCS...5...65G}), which are numerical techniques 
grounded in Bayesian statistics, to estimate the parameters of stars and determine which parameters align most 
effectively with the observed and synthetic spectra. In classical statistics, parameters are considered fixed 
values while observed data is treated as random variables. On the other hand, in Bayesian statistics, the observed 
data is regarded as fixed, with the objective being to estimate the parameters.

The code optimises up to eight stellar parameters each component at once: surface temperature ($T_{\rm eff\,1,2}$), surface 
gravity ($\log g_{1,2}$), metallicity ($[\rm M/H]_{1,2}$), limb darkening coefficients ($LD_{1,2}$ ) micro-turbulent 
velocity ($\nu_{mic,1{},2}$), macro-turbulent velocity ($\nu_{mac,1{},2}$), projected rotational velocity ($v \sin i_{1,2}$), and the fractional light contribution of the components to the total light ($l_{frac_{1,2}}$). The grid of theoretical spectra 
is constructed from every possible combination of these parameters, and the code evaluates each grid element concerning 
the observed normalised spectrum. The spectra are adjusted by the radial velocity of each component and then scaled to 
represent their contribution to the overall luminosity. The final result of this procedure is a composite spectrum. This 
synthetic composite spectrum can be generated to include the entire observed spectrum or limited to particular "segments" to 
reduce computing time and improve solutions. To determine the expected atmospheric parameters accurately, it is essential to 
use specific lines within the spectral range of the stars selected for atmospheric analysis. In this study, we neglected to 
make any specific distinction when determining the segments. We used the wavelength range between 4000-5200 $\AA$\,, where 
the spectral lines of both components are visible in the spectra.
\begin{table*}
\centering
\caption{Stellar atmospheric parameters are obtained from the methods described in detail in \S~\ref{spec_analysis}. The errors of the parameters in all tables of the present study are indicated in parentheses next to the values and represent the last digit(s).}
\label{tab:atmosphericParam}
\begin{tabular}{lcccccc}
\hline
\hline
Catalog~IDs & T$\rm_{eff}$ & $\log g$ & [M/H] & $v\sin i_{\rm obs}$ & \multirow{2}{*}{$l_{frac}$} & \multirow{2}{*}{Reduced $\chi^2$}		  \\
		    & (K)          & (dex)    & (dex) & (km~s$^{-1}$)       &            &   													  \\
\hline		    
\multirow{2}{*}{TIC\,426520557$^{\ddag}$}&28\,000(640)&3.812(45)&\multirow{2}{*}{-0.421(35)}&188(3)  &\multirow{2}{*}{1.0}  &\multirow{2}{*}{0.0051}\\
							       &---         &---      &                           &---     &	                  & 	                  \\
\multirow{2}{*}{TIC\,93997745}     &29\,700(610)&3.949(44)&\multirow{2}{*}{0.083(25)}&130(3)   &\multirow{2}{*}{0.683}&\multirow{2}{*}{0.0052}\\
								   &22\,000(640)&3.999(49)&                          &93(6)    &                      &                       \\		
\multirow{2}{*}{TIC\,339570153$^{\ddag}$}&31\,100(580)&4.259(59)&\multirow{2}{*}{0.109(47)}&182(3)   &\multirow{2}{*}{1.00} & \multirow{2}{*}{0.0053}\\
								   &---  	    &---  	  &		                     &---      &                      &                        \\
\multirow{2}{*}{TIC\,264540478}    &25\,600(610)&3.545(49)&\multirow{2}{*}{-0.172(38)}&102(3)  &\multirow{2}{*}{0.809}&\multirow{2}{*}{0.0049} \\
							       &21\,000(630)&3.791(51)&                           &77(4)   &                      &                        \\
\multirow{2}{*}{TIC\,379012185}&29\,900(610)&4.053(54)&\multirow{2}{*}{-0.263(34)} &78(4)      &\multirow{2}{*}{0.622}&\multirow{2}{*}{0.0048}\\
							   &19\,000(410)&3.572(51)&	                           &98(5)      &		              &                       \\
%
%\multirow{2}{*}{\red{TIC\,351701483$^*$}}&16\,500(610)&3.292(34)&\multirow{2}{*}{-0.577(24)}&263(9)  &\multirow{2}{*}{1.0}  &\multirow{2}{*}{0.0049}\\
%							       &---         &---      &	                          &---     &		              &                       \\
%
\hline
\end{tabular}\\
\footnotesize
$^{\ddag}$ During a total secondary eclipse, where the primary component occulted 99\% of the secondary ($\phi$ $\sim$ 0.5), the spectrum we analyse in this phase (as indicated in the brackets, Figure\,\ref{mcmc_tayf}) is purely linked with the primary component due to the total eclipse.
\end{table*}

In the solution, certain parameters are fixed (such as $\nu_{mic}$, $\nu_{mac}$, and $LD$), while relationships 
are established for others (for instance, $[\rm M/H]_1 = [\rm M/H]_2$ and $l_{frac_2} = 1 - l_{frac_1}$), thereby 
constraining the solution by these relationships. Furthermore, the variance ranges for independent parameters 
may be established, and the solution was confined to this range. The initial setup of the parameter seeking 
includes wide ranges with substantial step sizes to ensure the global minimum is identified while avoiding 
excessive computational time. In addition, separating macro-turbulent velocity from rotational velocity 
presents a challenge, necessitating spectra with extremely high resolution and a high SNR to address the 
degeneracy between these parameters \citep{2014MNRAS.444.3592D}. To prevent this degeneracy and to create 
a standard procedure for analysing spectra with lower resolution and SNR, the $\nu_{mic}$ and $\nu_{mac}$ 
are determined using the relations established by \citet{2010MNRAS.405.1907B} and \citet{2014MNRAS.444.3592D}. The 
research on the spectroscopic and asteroseismic data of main-sequence stars has produced these relationships, 
offering standard values of $\nu_{mic}$ (dependent on $T_{\rm eff}$) and $\nu_{mac}$ (dependent on $T_{\rm eff}$ 
and $\log g$) derived from a calibrated sample of main-sequence dwarfs. As a result, $\nu_{mic}$ and $\nu_{mac}$ 
are not adjusted freely in the subsequent code runs; rather, their values are determined for the next run 
according to the standard relations and the presently calculated $T_{\rm eff}$ and $\log g$.

A comprehensive atmospheric examination of the systems required the sampling of a substantial number of 
parameters. Determining how thoroughly the MCMC walkers explored the posterior distribution can be difficult, yet 
we can utilise certain heuristic measures to evaluate the advancement of these MCMC runs.

The atmospheric analysis of two systems (i.e., TIC\,426520557 and TIC\,339570153) is performed under the 
assumption of single-star spectra due to the existence of high signal-to-noise spectra at the secondary eclipse. Since 
these systems exhibit total eclipses, the spectra that are obtained at secondary eclipse are solely based on the primary 
component. Thus, the atmospheric analysis of these two systems was carried out with the assumption of a single star, while 
the remaining three systems were analysed as double-lined composite spectra.

The advantages and limitations of the strategies mentioned above are extensively analysed by \citet{2023MNRAS.526.5987C}. 
Figure\,\ref{mcmc_tayf} presents the results of our analysis of the samples, and for all the systems detailed in 
Table~\ref{tab:atmosphericParam}.
\begin{table*}
	\setlength{\tabcolsep}{2pt}
\renewcommand{\arraystretch}{1.0}
%\scriptsize
%\centering
\caption{The stellar parameters derived from light and radial velocity curve fit for all targets. The luminosity 
($L$) and bolometric magnitudes ($M_{\rm bol}$) were derived by adopting $T_{\rm eff}$ = 5\,780\,K and 
M$_{\rm bol}$ = +4.73 for solar values. }
\label{abs_para} 
\begin{tabular}{lccccc}
	\hline
	\hline
Parameter & TIC\,426520557      & TIC\,93997745       & TIC\,339570153      & TIC\,264540478      & TIC\,379012185      \\ 
		  & (primary/secondary) & (primary/secondary) & (primary/secondary) & (primary/secondary) & (primary/secondary) \\ 
\hline
T$_0$(BJD$-2,400,000$)$^a$&59478.4793(2)&58518.8242(1)&59363.6728(1)&58469.9122(1)&52856.736109(1)\\
$P$(day)&5.431750(3)&1.124480(1)&6.348762(4)&2.525940(2)&4.747168(2)\\
$a$ $sin\,i$ ($R_{\odot}$) &37.7(1)&4.8(1)&40.3(1)&21.9(1)&33.8(1)\\
$\gamma$(kms$^{-1}$)&$-$5.3(4)&27.1(7)&-51.2(5)&$-$11.8(1)&$-$45.1(3)\\
$K_{1,2}$ (km s$^{-1}$)&155.4(7) / 219.8(8) &83.3(6) / 131.6(6) & 67.8(3) / 253.7(7) & 174.3(1) / 265.6(3) & 115.1(1)/245.2(2)\\
$e$&0.0113(1)&0.0&0.0&0.0467(3)&0.0\\
$\omega$(${\degr}$)&163.5(6)&---&---&340.7(3)&---\\
$q$(M$_2$/M$_1$)&0.707(8)&0.63(1)&0.27(1)&0.66(2)&0.47(2)\\
$i$(${\degr}$)&78.1(5)&20.6(8)&89.8(2)&65.9(1)&62.1(1)\\
$T_{\rm eff\,1,2}$ (K)&28\,000$^b$ / 16\,300(420) &29\,700$^b$ / 23\,300(450) & 31\,100$^b$ / 17\,000(450) &25\,600$^b$ / 19\,900(420) & 29\,900$^b$/17\,500(160)\\
$\Omega_{1,2}$&7.9(1) / 7.7(1)&3.1(1) / 3.8(1)&6.9(1) / 5.5(1)&5.4(1) / 3.9(1)&3.6(1)/3.5(1)\\
$A_{1,2}$&1.0[fix]&1.0[fix]&1.0[fix]&1.0[fix]&1.0[fix]\\
$g_{1,2}$&1.0[fix]&1.0[fix]&1.0[fix]&1.0[fix]&1.0[fix]\\
$X_{1,2}^c$	&0.321 / 0.298 & 0.331 / 0.287 & 0.391 / 0.399& 0.289 / 0.291& 0.311/0.289\\
$l_{1}/(l_1 + l_2)$&0.81(4) & 0.83(3)&0.93(1)&0.58(1)&0.87(2)\\
$\ell_{3}$&---&---&0.248(11)&---&---\\
$r_{\rm 1,2}$&0.2136(1) / 0.1788(2)&0.4169(2) / 0.3566(3)& 0.1506(3) / 0.0647(2)&0.2098(1) / 0.2193(1) &0.3214(1)/0.2023(2)\\
$\Sigma$ $W(O-C)^2$	&0.002&0.003&0.001&0.002&0.001\\
\hline
Absolute parameters&\\
\hline
$M_{1,2}$ ($M_{\odot}$) &15.25(4) / 10.77(5)&16.30(3) / 10.32(4)&17.25(3) / 4.61(3)&17.69(5) / 11.61(5)&22.68(4) / 10.65(4)\\
$R_{1,2}$ ($R_{\odot}$) &8.23(4) / 6.89(4)  &5.66(2) / 4.84(2)  &6.07(1) / 2.61(2) &5.27(2) / 5.05(2)&12.29(3) / 7.74(5)\\
$\log(g_{1,2})$ (cgs)   &3.79(3) / 3.79(3)  &4.08(3) / 4.14(4)  &4.11(2) / 4.27(2) &4.06(3) / 4.28(4)&3.62(3) / 3.69(5)\\
$(v_{1,2}\sin i)_{\rm calc}$ (km s$^{-1}$)$^d$  &77(1) / 65(1) & 255(7) / 218(8)&48(1) / 21(1)&106(3) / 101(2)&130.9(1) / 82(1)\\
$\log(L_{1,2}/L_{\odot})$ &4.57(4) / 3.48(5)&4.36(4) / 3.79(4)&4.46(3) / 2.71(3)&3.59(3) / 3.99(4)&5.04(3) / 3.70(5)\\
$M_{\rm bol\,1,2}$ (mag) &$-$6.7(1) / $-$3.9(1)&$-$6.1(1) / $-$4.7(1)&$-$6.4(1) / $-$2.0(2)&$-$4.2(1) / $-$5.2(1)&$-$7.8(1)/$-$4.5(2)\\
$(m-M)_V$ (mag)    &13.9(6)&10.0(1)&12.3(1)&6.9(1)&13.5(1)\\
$E(B-V)$ (mag)      &0.85(2)&0.08(2)&0.43(1)&0.03(1)&0.288(11)\\
$d$ (pc)$^e$        &1\,786(55)&917(45)&1\,583(15)&351(4)&3\,336(61)\\
$d$ (pc)$^f$        &1\,815(101)&1\,055(125)&1\,570(47)&334(17)&3\,968(1\,357)\\
\hline
\end{tabular}\\
\scriptsize
$^a$Mid--time of the primary eclipse. $^b$T$_{\rm eff\,_1}$ were found in atmospheric 
analysis in \S~\ref{spec_analysis}. $^c$$X$ denotes linear coefficients of limb darkening. $^d$Theoretical 
projected rotational velocity calculated by \jktabs\ code under the assumption of (pseudo) synchronous 
rotation. $^e$The \jktabs\ distance is calculated from 2MASS 
magnitudes. $^f$From GAIA\footnote{\url{https://www.cosmos.esa.int/web/gaia/data-release-3}} trigomometric parallaxes.
\end{table*}

\section{Light curve modelling and absolute parameters}\label{lc}
To determine the physical characteristics of the systems, we used the fitting method, applying all \tes\ data 
for target stars, by using the Wilson-Devinney (hereafter WD) software \citep{WD_MAIN_1971ApJ}, developed in the 
\textsc{fortran} programming language. To effectively and rapidly employ the WD code, we utilise the \textsc{python} 
framework PyWD2015 \citep{PyWD2015_2020CoSka..50..535G}, which provides a graphical user interface for the 2015 version 
of the code \citep{WD2015_2014ApJ}. This routine allows for the simultaneous analysis of light and radial velocity 
curves. The WD program employs a differential corrections algorithm to identify a set of parameters that 
define observables. \texttt{Mode\,2} of the program was utilised because it was developed exclusively for 
detached eclipsing binaries. The albedos $A$ and gravity brightening $g$ coefficients for both components were set to 
be 1.0, which are standard values for stars having radiative envelopes \citep{1924MNRAS..84..702V,1969AcA....19..245R}. In 
all of our modelling, the linear limb darkening law was utilised with automatically updated coefficients from pre-computed 
tables provided by \citep{1993AJ....106.2096V}, along with the stellar atmosphere formulation. We established the grid 
integer size with \texttt{N1=N2=60} to define the dimensions of the grid at the stellar surface, and we set the noise 
parameter to \texttt{NOISE=1} to account for observational scatter that scales with the square root of the light level. The 
detailed settings aim to enhance the interpretation of our sample data.

The orbital periods and times of the primary eclipse minimum provided by \citet{2022ApJS..258...16P} were used 
to fold our light curves of the systems. The initial orbital parameters for the fitting were determined using 
the parameters from the \ravespan\ radial velocity fit (Section~\ref{ravespan}). We kept the effective temperatures 
of the primary component ($T_1$) constant throughout the light and radial velocity solution, assigning the temperature 
of the primary star based on the atmospheric analysis data obtained in the previous section. In essence, we assumed 
that the primary component is the hotter component and set $T_1$ as a fixed value. Similarly, we presumed 
that the metallicity corresponded to the estimated metallicity derived from the atmospheric analysis 
(see \S\,\ref{spec_analysis}).

\subsection{Modelling strategy}
The following parameters were modified simultaneously: the eccentricity $e$, the semi-major axis $a$, the orbital inclination 
$i$, the argument of periapsis $\omega$, the dimensionless surface potentials $\Omega_1$ and $\Omega_2$, the systemic velocity 
$V_{\gamma}$, the mass ratio $q$=$M_2$/$M_1$, the effective temperature of the secondary component $T_2$, the orbital period $P$, 
the phase shift $\varphi$, the light of the primary component in the specified filter $L_1$, and the third light $l_3$ as 
well. A comparison of the resulting $\chi$-squared value, which was derived by comparing the model to the observed light 
curves, was used to determine which solution was the most effective overall. In the method of light curve modelling, it 
became evident that the fractional radii of the components, $r_1$ and $r_2$, exhibit a significant correlation across all systems. We 
implemented the spectroscopic light ratio in the \tes\ band to resolve the degeneracy among the parameters. After reaching 
a satisfactory solution, we utilised new light ratios to reevaluate the effective temperature of the secondary components 
and adjusted the WD models with the updated temperatures. 

Although we found no signal for a tertiary component in the analysis of optical spectra, we tested all systems for the 
presence of a third light, $\ell_3$, which may originate from contaminated light of nearby sources. As we expected, we 
found a significant third light contribution in \tes\ light curves of TIC\,339570153. Since we do not detect any tertiary 
component in the radial velocity determination stage, we do not expect such a high third-light contribution. Therefore, the 
source of these contributions is very likely the light contamination from nearby objects to the target systems in the sky 
plane. However, we found insignificant values of $\ell_3$ for TIC\,379012185. Our analyses showed that adjusting $\ell_3$ 
in light curve analysis of these systems results in either very low or negative contribution, indicating that the \tes\ photometry of these sources did not significantly suffer from light contamination from nearby objects.

One of the stars examined is notable for its light curve. The early B-type star TIC\,93997745, believed 
to be a member of the IC\,2395 \citep{2003A&A...409..541C} galactic open cluster based on the 
GAIA\footnote{\url{https://www.cosmos.esa.int/web/gaia/data-release-3}} trigonometric parallaxes show no signs 
of eclipses and are likely an ellipsoidal variables. The ellipsoidal variations resulting from the non-spherical 
shapes of the components are particularly interesting as potential explanations for the out-of-eclipse modulation 
observed in the \tes\ light curve. High-resolution spectroscopy has identified TIC\,93997745 as SB2 system 
(see Figure\,\ref{mcmc_tayf}). Additionally, it displays uneven maxima in its \tes\ light curve, which may 
suggest the presence of pulsations.

\begin{figure*}
	\center
    \includegraphics[width=1\textwidth]{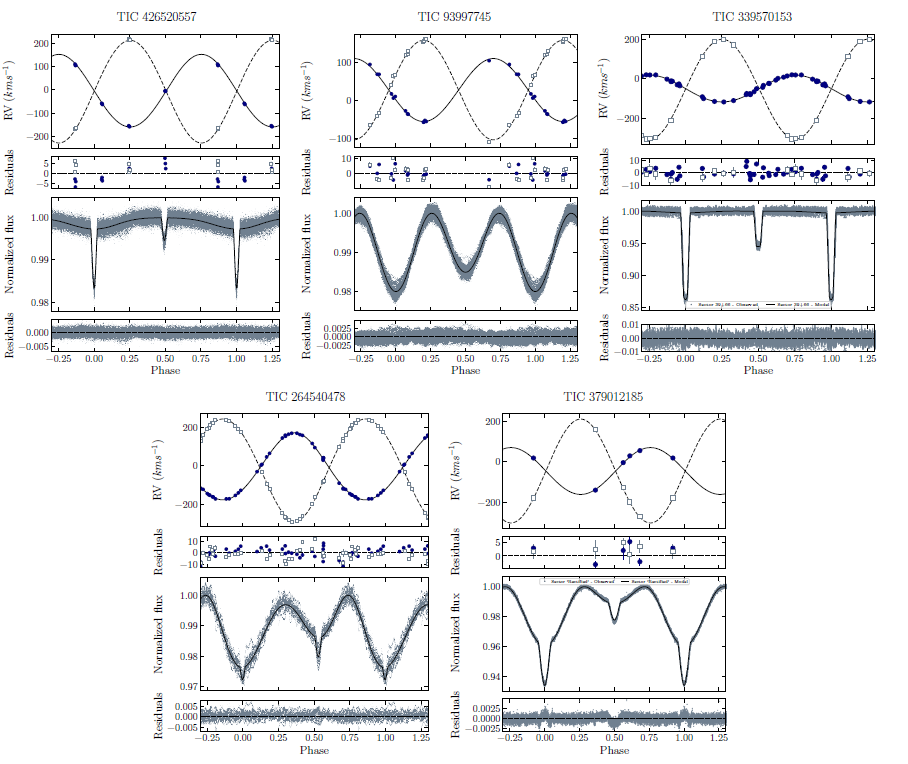}
\caption{The combination of both analysis of the \tes\ photometry and radial velocity time series of the five eclipsing binaries. The \tes\ light curve for each star is displayed in the lower two 
panels. Continuous curves represent the models that exhibit the fit. The radial velocity plot clearly distinguishes between the primary and secondary stars, with filled symbols representing the primary and open symbols representing the secondary. This distinction is evident in the upper two panels for each star. The most likely models for the primary and secondary stars are represented 
by solid and dashed lines, respectively.
}
\label{LC_RV_plot}
\end{figure*}

Upon achieving the best solution, we consider all parameters as adjustable and perform a single iteration to assess 
statistical uncertainties, alongside the results presented in Table~\ref{abs_para}. We do not provide the internal 
error of $T_2$ due to its implausibly small value (a few K); thus, we utilise the uncertainty of $T_2$ determined 
in \S\,\ref{spec_analysis} as its uncertainty. Presumably, internal errors of spectroscopic parameters printed in 
Table~\ref{abs_para} is driven by the scatter of the residual radial velocities. This may prevent estimation of more realistic 
uncertainties, particularly for systems with a smaller number of observations (e.g., TIC 426520557 and TIC 379012185). We 
estimate that the external errors of spectroscopic parameters are likely four to six times larger than the formal 
errors printed in Table~\ref{abs_para}. Figure~\ref{LC_RV_plot} presents the optimal models of solutions for the 
light and radial velocity curves, with the residuals from both solutions displayed in the bottom panel within the 
same frame.
%

%
%
%
%%%%%%%%%%%%%%%%%%%%%%%%%%%%%%%%%%%%%%%%%%%%%%%%%%%%%%%%%
% Figure 04
%%%%%%%%%%%%%%%%%%%%%%%%%%%%%%%%%%%%%%%%%%%%%%%%%%%%%%%%%%
%
%
%

\subsection{Absolute parameters and uncertainties}
Prior to comparing the model predictions with the observational data, we established all of the orbital and stellar 
properties of the systems. The existing literature provides limited information concerning the absolute stellar 
parameters of the systems. Considering this, we determined that, regardless of the existing literature, it 
would be most effective to analyse each of them uniformly, utilising the results from the previous sections 
on photometry and spectroscopy. This was achieved through the application of the \jktabs\ code 
\citep{2005A&A...429..645S}, which employs perturbation analysis to address uncertainty effectively. This 
obvious procedure necessitates the orbital period, eccentricity, fractional radii, velocity semi-amplitudes, and 
inclination as inputs, each associated with statistical uncertainties, and yields the absolute values of masses 
and radii (in specified units) as outputs. In analysing solar units, $\log(g)$, rotational velocities, and 
synchronisation, we assume tidal locking is in effect. \jktabs\ calculates the distance to an object by 
using the effective temperatures of two components, the approximate metallicity (with a precision of 0.5\,dex), 
$E$(\B-\V) color excess, and the apparent magnitudes. The code does not function with brightness in the \tes\ band. 

The code utilises the apparent total magnitudes of an individual binary across the \U, \B, \V, \R, \I, \J, $H$, 
and \K\ bands to achieve this. This approach evaluates the total magnitudes observed against the absolute magnitudes 
determined through various bolometric corrections \citep{1976ApJ...203..417C, 1996ApJ...469..355F,1998A&A...333..231B,
2002A&A...391..195G} and the surface brightness-$T_{\rm eff}$ relationships established by \citet{2004A&A...426..297K}. 
Furthermore, flux ratios can be employed to compute the absolute magnitudes of each component. The final distance 
value is determined by a weighted average of five values, which are calculated for each band based on the surface 
brightness-$T_{\rm eff}$ relations from \citet{2004A&A...426..297K}. The results can be compared with trigonometric parallaxes 
from the $Gaia$ DR3 \citep{2023A&A...674A...1G}. Together with stellar, photometric, and orbital parameters, \jktabs\ computes 
expected rotation velocities for the scenario of synchronisation of rotation with the orbital period $v_{syn}$, in 
addition to the time scale for this synchronisation and the time scale for the circularisation of the orbit. 

The results derived from the code are displayed in Table~\ref{abs_para}. The resulting parameters have been used 
for positioning the components on the HR diagrams to analyse their evolutionary status.
%

%
%
%
%%%%%%%%%%%%%%%%%%%%%%%%%%%%%%%%%%%%%%%%%%%%%%%%%%%%%%%%%
%Figure 05
\begin{figure*}
	\center
\includegraphics[width=1\textwidth]{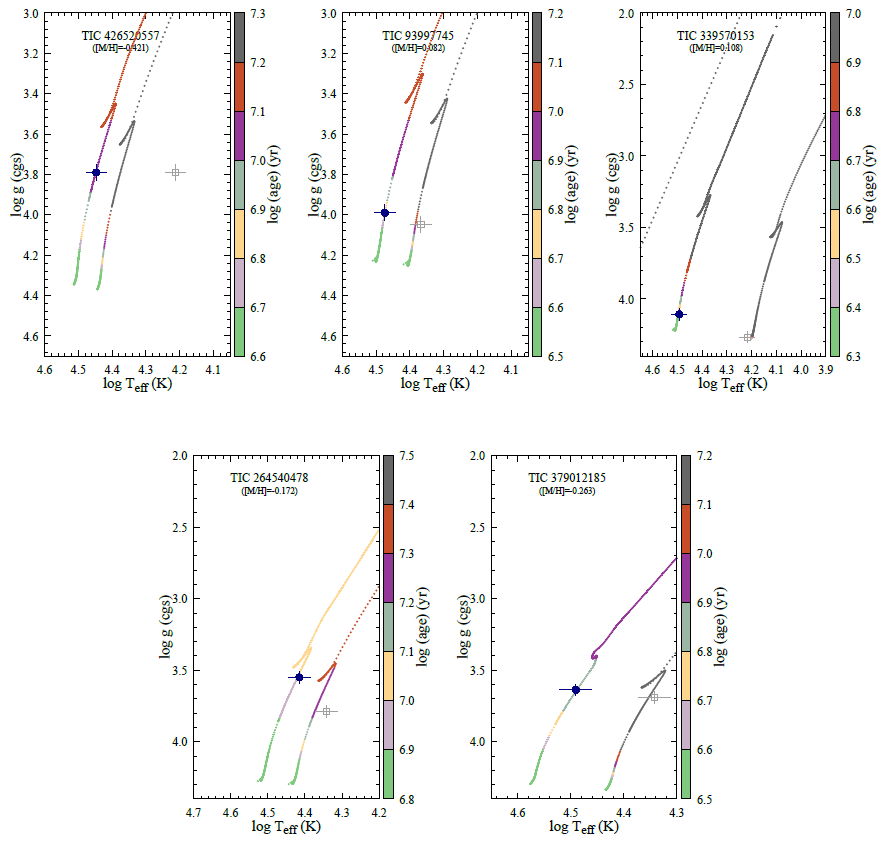}

\caption{Comparison of our results with MESA evolution tracks on T$_{\rm eff}$/$\log(g)$ planes. The primary components are shown with the filled circles, whilst the secondaries are represented with the open squares, both with error bars from Table\,\ref{abs_para}. The coloured points are the evolutionary tracks that best match the calculated component masses, and the colours indicate the stellar age with the colour bars at the right side of the plots.}
\label{HR}
\end{figure*}
%%%%%%%%%%%%%%%%%%%%%%%%%%%%%%%%%%%%%%%%%%%%%%%%%%%%%%%%%%
%
%
%

\section{Evolutionary status}\label{sec_iso}
As mentioned in \S\,\ref{intro}, various researchers calculated the instability region for $\beta$\,Cephei stars within the HR diagram. The $\beta$\,Cephei pulsators are typically classified as early-type B stars, exhibiting light and radial velocity variations over periods of several hours. Accurately classifying 
B-type variable stars prove to be challenging, especially due to the overlap that exists among the various groups of variable stars \citep{1996A&A...315L.401H}. The boundaries established by these stars rely on absolute parameters such as mass, temperature, and radius, which are obtained from light and radial velocity 
analysis, along with atmospheric parameters determined through precise methods. Plotting the stars on the HR diagram provides us a clear understanding of the precise position of the $\beta$\,Cephei pulsating systems within the diagram.

We determined the absolute parameters of the $\beta$\,Cephei stars examined in this study to enhance our understanding and make it easier to compare with theoretical results, thereby enabling their placement in the HR diagram. In comparison with theoretical models, we also employed the stellar evolution models. We use the \textsc{mesa} Isochrones and Stellar Tracks \citep[MIST\footnote{\url{https://waps.cfa.harvard.edu/MIST}} v1.2,][]{2016ApJS..222....8D, 2016ApJ...823..102C}, which is based on the Modules for Experiments in Stellar Astrophysics (\textsc{mesa\footnote{\url{https://mesastar.org/}}}) package 
\citep{2011ApJS..192....3P,2013ApJS..208....4P,2015ApJS..220...15P} and offers evolutionary tracks for a wide range of stars by employing various metallicities.

High-mass star evolutionary models are frequently challenged to match up with observational data precisely. Mass is the key stellar quantity that determines the path of stellar evolution, and it can be measured in a model-independent way within binary systems. Figure\,\ref{HR} presents the positions of the components in HR diagrams. The colour bar on the right shows the theoretical evolutionary tracks corresponding to their respective age values, visually representing various age ranges to facilitate the interpretation of the evolutionary path. The theoretical models match the primaries of all five systems. The secondary of the 
TIC\,426520557 is notably far away from its theoretical predictions. It contributes significantly to the total flux; however, it is felt as very weak in the SOPHIE spectra of this system. The insufficient amount of measured radial velocities for the secondary component could lead to an inaccurate estimation of its mass. As a result, this influences the alignment with the evolutionary path established by the component's determined mass. This problem requires additional research via scheduled high-resolution spectral observations. Furthermore, in all systems studied, it seems that the secondary components of the systems typically consumed their central hydrogen more rapidly. This conclusion suggests 
that a mass transfer process may have taken place earlier in these systems.
%
%
%
%
%%%%%%%%%%%%%%%%%%%%%%%%%%%%%%%%%%%%%%%%%%%%%%%%%%%%%%%%%
%Figure 06
\begin{figure*}
\center
    \includegraphics[width=1\textwidth]{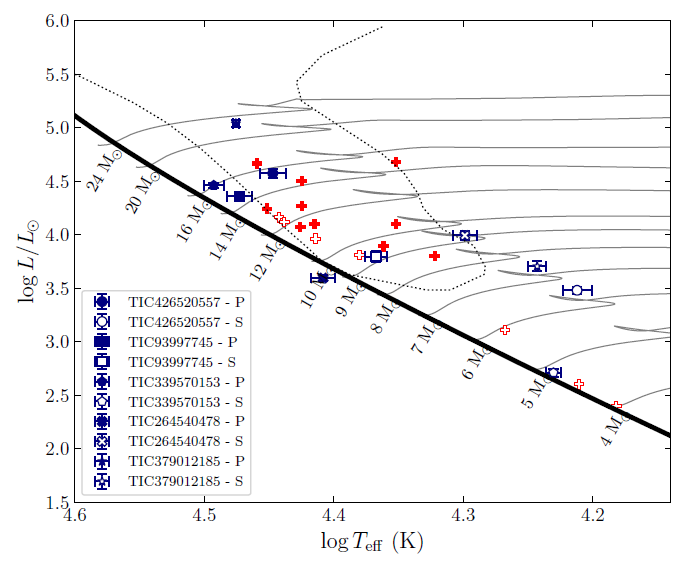}
\caption{The theoretical HR diagram of the confirmed $\beta$\,Cephei stars are presented in this work, along with 
references from the literature \citep{2021AJ....161...32L,2021MNRAS.501L..65S, 2022MNRAS.513.3191S,2020MNRAS.497L..19S,2025A&A...693A.103S}, marked by plus symbols. The solid line indicates the ZAMS \citep{1996MNRAS.281..257T}, whereas the thick dashed line marks the limits of the theoretical $\beta$\,Cephei instability strip \citep{1999AcA....49..119P}. We illustrate various stellar evolutionary tracks (thin curves), each identified by its corresponding evolutionary masses.}
\label{HR_beta_cep}
\end{figure*}
%%%%%%%%%%%%%%%%%%%%%%%%%%%%%%%%%%%%%%%%%%%%%%%%%%%%%%%%%%
%
%
%

Figure\,\ref{HR_beta_cep} illustrates the confirmed $\beta$\,Cephei-type stars found in eclipsing binaries, identified through photometric and spectroscopic techniques. The theoretical boundaries of the $\beta$\,Cephei instability strip were adopted from the work of \citet{1999AcA....49..119P}. $\beta$\,Cephei type variables in eclipsing binary systems have been uncommon, as \citet{2024ApJS..272...25E} have recently reported the discovery of 78 pulsators of the $\beta$\,Cephei type in eclipsing binaries using \tes\ with 59 being new findings. However, there have been no spectral observations to establish absolute parameters for placement in the HR diagram. The stars are located in a particular region on the HR diagram, except for our samples, which appear to fall outside the designated lines proposed by \citet{1999AcA....49..119P}. The confirmed $\beta$\,Cephei-type stars in eclipsing binaries are, in contrast, widely scattered. The theoretically predicted instability strip is acknowledged as not covering a complete set of stars, as emphasised in the significant discussions by \citet{2005ApJS..158..193S} and \citet{2024ApJS..272...25E}, which take binary systems into account. Additionally, the mass distribution of the $\beta$\,Cephei stars is noteworthy in the strip. The confirmed $\beta$\,Cephei stars in eclipsing binaries display a mass distribution that sharply peaks at approximately 12\,$M_{\odot}$ in Stankov's diagram and 15\,$M_{\odot}$ in Eze's HR diagram, which also delineates the boundary for O-type stars. This distribution is significant due to the diminishing presence of pulsating stars among more massive stars. The empirical identification of the boundaries of the instability region is a compelling problem in the study of $\beta$\,Cephei-type stars. Therefore, recent surveys of late O-type stars and early to mid B-type stars would hold significant astrophysical relevance.
%
%
%
%
%%%%%%%%%%%%%%%%%%%%%%%%%%%%%%%%%%%%%%%%%%%%%%%%%%%%%%%%%
%Figure 07
\begin{figure*}
\center
    \includegraphics[width=0.97\textwidth]{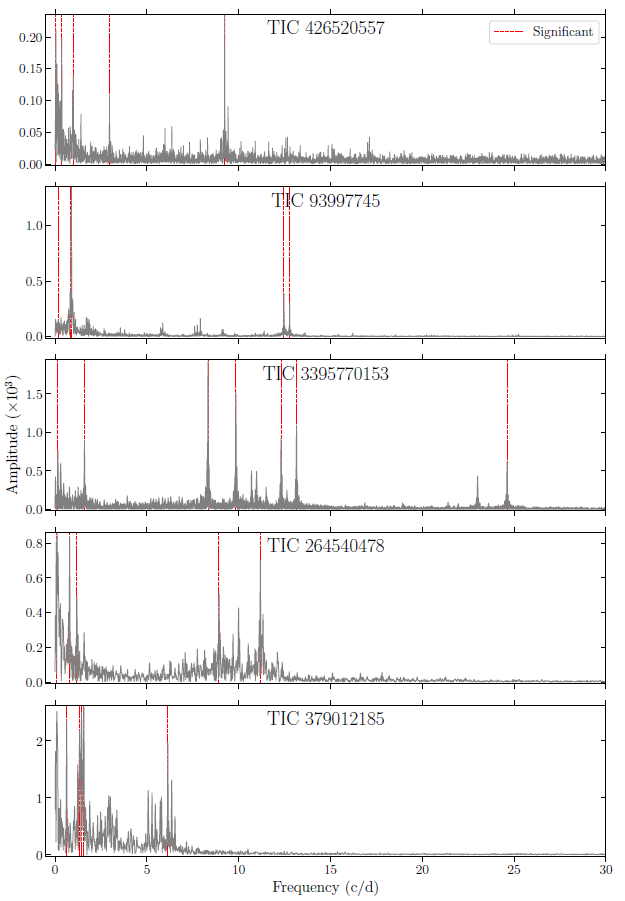}
\caption{The summary figures are intended for systems that demonstrate considerable relative pulsation 
amplitudes. The panel presents the amplitude spectrum corresponding to the residual light curve, featuring 
significant (genuine) pulsation mode frequencies marked by vertical dashed lines.
}
\label{puls}
\end{figure*}
%%%%%%%%%%%%%%%%%%%%%%%%%%%%%%%%%%%%%%%%%%%%%%%%%%%%%%%%%%
%
%
%

\section{Pulsation}\label{pulsi}
Following the removal of the binary model, we used the residual light curve for each of the systems to obtain 
an amplitude spectrum using a discrete Fourier transform \citep{1985MNRAS.213..773K}. All residuals 
of combined \tes\ light curves from the target stars were used in the Fourier analysis. In a method similar to
\citet{2023MNRAS.523.1676C}, we accepted significant and resolved pulsation frequencies (following 1/(T), where 
T represents the duration of the \tes\ dataset) and determined their optimised parameters through a non-linear 
least-squares cosinusoid fit applied to the residual light curve utilising \sigspec\ software 
\citep{2007A&A...467.1353R}. We established significance values higher than 5 as our significance criterion, in 
line with prior research on \tes\ data that recommends values exceeding the common SNR=4 threshold set by 
\citet{1993Ap&SS.210..173B} \citep[see e.g.][]{2021AcA....71..113B}.

A summary Figure\,\ref{puls} was generated for each frequency analysis. In the systems that underwent eclipse 
removal before frequency analysis, the significant frequencies are indicated with vertical dashed lines and 
illustrated in the figure. We also present the genuine pulsation mode frequencies, amplitudes, and phases, along 
with their uncertainties, for each star as derived from the non-linear least-squares fit in Table\,\ref{tab:Freqs}. The 
errors of all values listed in Table\,\ref{tab:Freqs} are derived according to the method described in 
\citet{2008A&A...481..571K}.

We must point out that we observed a notable issue with stellar pulsation in this instance, where, in the 
context of pulsating binaries, there is uncertainty regarding which star in the system is exhibiting 
pulsation. Unfortunately, the \tes\ data by themselves do not enable us to certainly determine which 
star is pulsating. We do not observe amplitude modulation of the pulsation modes throughout the 
orbit, indicating that nearly all systems are not single-sided pulsators \citep{2020NatAs...4..684H}.

Systems TIC\,426520557, TIC\,339570153, and TIC\,379012185 were previously classified as $\beta$\,Cephei 
by \citet{2024ApJS..272...25E}; however, a comprehensive frequency analysis has not been conducted. The absence 
of analysis has created a notable deficiency in our comprehension of their pulsation characteristics. This study 
presents analyses of light and radial velocity curves, as well as atmospheric analysis, for the first time by 
including data from later sectors of \tes. This study additionally performed light and radial velocity curve 
analyses for the rest of the systems for the first time, together with atmospheric analyses. Frequency analyses 
of the $\beta$\,Cephei candidates were also performed.

The combination of theoretical understandings of asteroseismology with empirical determinations of fundamental 
stellar parameters in $\beta$\,Cephei stars within eclipsing binaries presents a significant opportunity for 
comprehensive asteroseismic analysis of these pulsators. This approach holds significant potential for addressing the complex and elusive discrepancies observed between stellar structure models and the evolution of massive stars 
with pulsating components.

The HR diagram includes several groups of pulsating stars positioned within the theoretical pulsational instability 
strip \citep{2004A&A...414L..17D,2017MNRAS.469...13S}. The studies by \citet{1999AcA....49..119P} and 
\citet{2001MNRAS.327..881D} clarified the theoretical limits of the $\beta$\,Cephei instability strip by employing 
enhanced opacity tables and more accurate interpolation methods. Another study that deserves attention is 
\citet{2007MNRAS.375L..21M} which examined how uncertainties in opacity calculations and the solar metal mixture 
influence the excitation of pulsation modes in B-type stars, proposing that the newly identified pulsational 
stabilities in HR diagram correspond to unstable domains. However, the limited presence of $\beta$\,Cephei-type 
pulsations in eclipsing binary stars has led to an absence of significant discoveries thus far. As a result, the 
HR diagram indicates that, aside from a small number of samples \citep{2008A&A...477..917P,2024CoSka..54b..70E}, 
O-type $\beta$\,Cephei stars do not enhance our knowledge of their mass distribution. Therefore, this study of 
massive binaries containing $\beta$\,Cephei pulsators provides an opportunity to explore further the complexity 
of pulsation modes and their underlying mechanisms.

We found that a considerable number of frequencies are very likely to grow to observable amplitudes. The values 
presented in the  Table\,\ref{tab:Freqs} represent solely the genuine frequencies identified; they do not include all 
the peaks detected in the frequency spectrum. The frequency spectrum reveals additional peaks that are harmonics and 
combinations of the genuine frequencies, and therefore, they are not listed in the  Table\,\ref{tab:Freqs}.

\section{Conclusions}
\label{conc}
Despite significant efforts in theory and observation, our understanding of the structure and evolution of 
$\beta$\,Cephei pulsators in eclipsing binaries remain incomplete. Our study aims to enhance the quantity of eclipsing 
binaries possessing $\beta$\,Cephei pulsators and to obtain accurate measurements of their physical properties, thereby 
creating a statistically significant sample that can assist in evaluating the predictions of theoretical models. To this 
aim, we study five binary systems to the list, featuring stars with masses varying from 4.6 to 22.6\,$M_{\odot}$. The 
spectroscopic orbits and atmospheric analysis of these systems are examined, with the full physical properties of the 
systems calculated based on the high-resolution spectroscopy and \tes\ photometry. All of the objects examined in this 
work exhibit lovely eclipses and pulsations, making them ideal candidates for further investigation.

\section{acknowledgements}
We thank the anonymous reviewer for their careful reading of our work and their many insightful 
comments and suggestions. This study was funded by Scientific Research Projects Coordination Unit 
of Istanbul University. Project number: FBA-2024-40612. This paper used the following internet-based 
resources for research: the NASA Astrophysics Data System and the SIMBAD database managed by CDS at 
Strasbourg Observatory, France. The photometric data was acquired from the Mikulski Archive for 
Space Telescopes (MAST). This study employs data from the European Space Agency (ESA) mission 
\gaia\, which was processed by the \gaia\ Data Processing and Analysis Consortium.

\section*{Data Availability}
Photometric and spectroscopic raw data used in this paper are publicly available at the
\url{https://archive.stsci.edu/missions-and-data/tess}, and \url{http://archive.eso.org/cms.html} 
archives.

\bibliographystyle{mnras}
\bibliography{Bibliography}

%%%%%%%%%%%%%%%%% APPENDICES %%%%%%%%%%%%%%%%%%%%%

\appendix
\section{Radial velocity measurements of targets}
\fontsize{9}{11}\selectfont
%\centering
\onecolumn
\begin{longtable}{l c c c c c c c c}
	\caption{Log of the radial velocity measurements of the targets.}
	\label{tab:target_RVs}\\
	\toprule
	%\hline
	\makecell[b]{System\\~}
	& \makecell[b]{HJD\\(+2400000)}
	& \makecell[b]{$v_1$\\(\kms)}
	& \makecell[b]{$\sigma_1$\\(\kms)}
	& \makecell[b]{$v_2$\\(\kms)}
	& \makecell[b]{$\sigma_2$\\(\kms)}
	& \makecell[b]{S/N$^a$\\~}
	& \makecell[b]{Instrument\\~}\\
	
	\midrule
	%\hline
	\endfirsthead
	\caption{(Cont.)}\\
	\toprule
	%\hline
	
	\makecell[b]{System\\~}
	&\makecell[b]{HJD\\(+2400000)}
	& \makecell[b]{$v_1$\\(\kms)}
	& \makecell[b]{$\sigma_1$\\(\kms)}
	& \makecell[b]{$v_2$\\(\kms)}
	& \makecell[b]{$\sigma_2$\\(\kms)}
	& \makecell[b]{S/N$^a$\\~}
	& \makecell[b]{Instrument\\~}\\
	\midrule
	%\hline
	\endhead
	\midrule
	%\hline
	\multicolumn{8}{r}{\footnotesize\itshape Continue on the next page}
	\endfoot
	\bottomrule
	%\hline
%	       &      &       &       &       & TIC\,351701483  
	\multicolumn{8}{l}{\footnotesize $^a$ S/N values have been obtained from headers.}
	\endlastfoot
TIC\,426520557&60290.51290 &   -5.1 & 0.2 &  ---  &  --- & 35 & SOPHIE\\
&60290.52778 &   -5.2 & 0.2 &  ---  &  --- & 36 & SOPHIE\\
&60290.54263 &   -5.1 & 0.2 &  ---  &  --- & 32 & SOPHIE\\
&60292.51691 &  110.0 & 0.2 & -165.9&  0.2 & 76 & SOPHIE\\
&60292.53197 &  104.1 & 0.2 & -169.1&  0.2 & 67 & SOPHIE\\
&60292.54700 &  105.0 & 0.2 & -162.8&  0.2 & 59 & SOPHIE\\
&60293.53421 &  -58.0 & 0.2 &  ---  &  --- & 59 & SOPHIE\\
&60293.54688 &  -59.1 & 0.2 &  ---  &  --- & 59 & SOPHIE\\
&60293.55942 &  -62.9 & 0.2 &  ---  &  --- & 58 & SOPHIE\\
&60294.56595 & -153.7 & 0.2 & 216.4 &  0.2 & 72 & SOPHIE\\
&60294.58082 & -156.2 & 0.2 & 212.3 &  0.2 & 71 & SOPHIE\\
&60294.59576 & -158.0 & 0.2 & 213.7 &  0.2 & 69 & SOPHIE\\
\midrule
TIC\,93997745&53067.51934 &  -57.9 & 0.2 &  154.8 &  0.7 & 293 & HARPS\\
&53067.52373 &  -57.3 & 0.2 &  156.2 &  0.5 & 292 & HARPS\\
&53067.53122 &  -53.2 & 0.2 &  161.2 &  0.5 & 244 & HARPS\\
&53067.54215 &  -56.0 & 0.2 &  161.2 &  0.3 & 223 & HARPS\\
&53068.49948 &  -28.0 & 0.2 &  120.2 &  2.2 & 204 & HARPS\\
&53068.52075 &  -36.8 & 0.2 &  130.2 &  2.2 & 208 & HARPS\\
&53069.51070 &   16.9 & 0.2 &   40.2 &  0.2 & 249 & HARPS\\
&53069.52572 &    5.9 & 0.2 &   64.2 &  0.2 & 242 & HARPS\\
&53069.54085 &   10.0 & 0.2 &   67.5 &  1.2 & 240 & HARPS\\
&53070.52014 &   68.9 & 0.2 &  -41.8 &  0.2 & 229 & HARPS\\
&53070.53516 &   68.7 & 0.2 &  -32.8 &  0.2 & 239 & HARPS\\
&53291.82254 &  104.9 & 0.2 & -110.8 &  0.2 & 317 & UVES \\
&53371.82459 &   94.9 & 0.2 &  -65.8 &  1.2 & 282 & UVES \\
&53430.60125 &  -36.0 & 0.2 &  121.2 &  0.2 & 310 & HARPS\\
\midrule
TIC\,339570153&	57845.76590	&	-15.4	&	1.2	&	-211.2	&	3.2	&	431	&	GIRAFFE	\\
&	57919.64630	&	-34.4	&	1.1	&	---	&	---	&	361	&	GIRAFFE	\\
&	57920.68290	&	13.5	&	0.5	&-288.7	&	2.5	&	976	&	GIRAFFE	\\
&	57926.53400	&	-7.7	&	1.2	&	---	&	---	&	542	&	GIRAFFE	\\
&	57929.68360	&	-92.2	&	0.6	&111.6	&	3.4	&	416	&	GIRAFFE	\\
&	57932.57590	&	-24.1	&	1.2	&	---	&	---	&	626	&	GIRAFFE	\\
&	57934.72540	&	-13.4	&	1.3	&	---	&	---	&	648	&	GIRAFFE	\\
&	57946.70010	&	18.7	&	1.2	&-299.8	&	4.6	&	636	&	GIRAFFE	\\
&	57947.63580	&	-31.7	&	2.1	&	---	&	---	&	612	&	GIRAFFE	\\
&	57950.64310	&	-81.2	&	1.1	&	---	&	---	&	539	&	GIRAFFE	\\
&	57955.61300	&	-115.9	&	0.8	&188.2	&	2.2	&	526	&	GIRAFFE	\\
&	57957.71290	&	-44.2	&	1.3	&	---	&	---	&	611	&	GIRAFFE	\\
&	57963.50990	&	-77.3	&	1.1	&	---	&	---	&	661	&	GIRAFFE	\\
&	57971.68370	&	19.3	&	0.9	&-305.7	&	1.5	&	664	&	GIRAFFE	\\
&	57982.57390	&	-78.2	&	1.1	&	---	&	---	&	615	&	GIRAFFE	\\
&	57998.55210	&	-36.6	&	0.5	&	---	&	---	&	497	&	GIRAFFE	\\
&	58001.58940	&	-78.4	&	0.9	&	---	&	---	&	547	&	GIRAFFE	\\
&	58001.62520	&	-81.2	&	1.1	&	---	&	---	&	478	&	GIRAFFE	\\
&	58003.59360	&	18.4	&	0.7	&-303.8	&	3.3	&	580	&	GIRAFFE	\\
&	58004.51590	&	-11.4	&	0.8	&	---	&	---	&	763	&	GIRAFFE	\\
&	58014.48370	&	-66.9	&	0.9	&	---	&	---	&	697	&	GIRAFFE	\\
&	58015.52490	&	1.2	    &	0.8	&	---	&	---	&	564	&	GIRAFFE	\\
&	58017.56050	&	-26.9	&	0.9	&	---	&	---	&	583	&	GIRAFFE	\\
&	58019.51580	&	-118.5	&	0.9	&198.7	&	2.2	&	514	&	GIRAFFE	\\
&	58020.51660	&	-75.6	&	1.2	&	---	&	---	&	709	&	GIRAFFE	\\
&	58021.49590	&	-24.7	&	0.7	&	---	&	---	&	628	&	GIRAFFE	\\
&	58221.82670	&	-101.6	&	1.1	&	---	&	---	&	854	&	GIRAFFE	\\
&	58236.83640	&	-51.2	&	1.4	&	---	&	---	&	536	&	GIRAFFE	\\
&	58250.71570	&	7.3	    &	0.9	&	---	&	---	&	465	&	GIRAFFE	\\
&	58264.67880	&	-2.7	&	0.7	&	---	&	---	&	625	&	GIRAFFE	\\
&	58286.63230	&	-111.5	&	1.1	&171.2	&	4.4	&	757	&	GIRAFFE	\\
\midrule
TIC\,264540478&58013.83710 &   27.7 &  0.2 &  -86.6 & 0.2 & 490 & UVES\\
&58013.83716 &   31.8 &  0.2 &  -85.6 & 0.2 & 484 & UVES\\
&58013.83721 &   38.4 &  0.2 &  -83.6 & 0.2 & 365 & UVES\\
&58013.84349 &   38.6 &  0.2 &  -81.6 & 0.2 & 409 & UVES\\
&58013.84366 &   40.8 &  0.2 &  -81.6 & 0.2 & 447 & UVES\\
&54147.37656 &    6.4 &  0.2 &  -34.2 & 0.2 & 316 & SOPHIE\\
&51141.39068 &  152.6 &  0.1 & -272.6 & 0.4 & 390 & ELODIE\\
&51141.44231 &  162.7 &  0.1 & -282.3 & 0.4 & 472 & ELODIE\\
&51141.49871 &  167.9 &  0.1 & -291.4 & 0.4 & 396 & ELODIE\\
&51141.59160 &  168.2 &  0.1 & -286.5 & 0.4 & 391 & ELODIE\\
&51141.64503 &  160.8 &  0.1 & -278.8 & 0.4 & 253 & ELODIE\\
&51141.69034 &  156.8 &  0.1 & -262.0 & 0.4 & 338 & ELODIE\\
&51142.40984 & -121.5 &  0.1 &  137.5 & 3.0 & 376 & ELODIE\\
&51142.45470 & -123.9 &  0.1 &  159.1 & 0.5 & 354 & ELODIE\\
&51142.55646 & -152.2 &  0.1 &  194.8 & 0.5 & 407 & ELODIE\\
&51142.62053 & -162.5 &  0.1 &  220.8 & 0.5 & 300 & ELODIE\\
&51142.67381 & -173.4 &  0.1 &  234.5 & 2.2 & 346 & ELODIE\\
&51143.41118 &  -31.8 &  0.1 &  ---   & --- & 201 & ELODIE\\
&51143.48398 &    2.2 &  0.1 &  ---   & --- & 233 & ELODIE\\
&51143.58025 &   44.8 &  0.2 &  -94.4 & 0.2 & 106 & ELODIE\\
&51143.63282 &   63.3 &  0.2 & -124.1 & 0.5 & 297 & ELODIE\\
&51144.38950 &  114.7 &  0.1 & -198.9 & 0.4 & 339 & ELODIE\\
&51144.43564 &   93.2 &  0.2 & -164.3 & 0.5 & 148 & ELODIE\\
&51145.39160 & -173.4 &  0.1 &  235.8 & 0.4 & 210 & ELODIE\\
&51145.44370 & -172.9 &  0.1 &  231.5 & 0.4 & 364 & ELODIE\\
&51145.55291 & -155.2 &  0.1 &  207.7 & 0.4 & 391 & ELODIE\\
&51145.60555 & -142.2 &  0.1 &  191.2 & 0.4 & 324 & ELODIE\\
&51145.64852 & -128.8 &  0.1 &  175.4 & 0.5 & 293 & ELODIE\\
&51146.38625 &  143.2 &  0.1 & -244.3 & 0.5 & 176 & ELODIE\\
&51146.43632 &  157.8 &  0.1 & -268.3 & 0.4 & 335 & ELODIE\\
&51147.39220 &  -91.6 &  0.2 &   95.4 & 4.8 & 312 & ELODIE\\
&51147.46455 & -118.5 &  0.2 &  130.1 & 0.7 & 340 & ELODIE\\
&51147.58181 & -145.9 &  0.1 &  188.0 & 0.5 & 390 & ELODIE\\
&51147.62574 & -153.9 &  0.1 &  210.7 & 0.5 & 278 & ELODIE\\
&51147.68657 & -163.2 &  0.1 &  221.1 & 2.0 & 278 & ELODIE\\
\midrule
TIC\,379012185&	54600.65242	&	-138.6	&	1.1	&	158.851	&	3.6	&	228	&	FEROS	\\
&	54601.59915	&	-3.5	&	3.5	&	-126.113&	1.5	&	231	&	FEROS	\\
&	55642.88320	&	19.2	&	1.1	&	-177.804&	2.6	&	168	&	FEROS	\\
&	55698.72392	&	55.6	&	1.3	&	-268.28	&	2.4	&	181	&	FEROS	\\
&	56068.64615	&	29.8	&	1.6	&	-197.09	&	3.4	&	209	&	FEROS	
\label{tab:RVs}
\end{longtable}
	%%%%%%%%%%%%%%%%% APPENDICES %%%%%%%%%%%%%%%%%%%%%

	\section{Frequencies}
\fontsize{7}{11}\selectfont
%\centering
\onecolumn
\begin{longtable}{c c c c c c}
	\caption{Genuine frequencies found in the light curve residuals of the targets.}
	\label{tab:Freqs}\\
	\toprule
	%\hline
	  \makecell[b]{ID\\~}
	& \makecell[b]{Frequency\\($c/d$)}
	& \makecell[b]{$Amplitude$\\($\times10^{-3}$)}
	& \makecell[b]{Phase\\(rad)}
	& \makecell[b]{Significance\\~}
	& \makecell[b]{RMS\\~}\\
	% & \makecell[b]{Combination\\~}\\

	\midrule
	%\hline
	\endfirsthead
	\caption{(Cont.)}\\
	\toprule
	%\hline

	  \makecell[b]{ID\\~}
	& \makecell[b]{Frequency\\($c/d$)}
	& \makecell[b]{$Amplitude$\\($\times10^{-3}$)}
	& \makecell[b]{Phase\\(rad)}
	& \makecell[b]{Significance\\~}
	& \makecell[b]{RMS\\~}\\
	% & \makecell[b]{Combination\\~}\\
	\midrule
	%\hline
	\endhead
	\midrule
	%\hline
	\multicolumn{5}{r}{\footnotesize\itshape Continue on the next page}
	\endfoot
	\bottomrule
	%\hline
	\endlastfoot
 \multicolumn{5}{l}{TIC 426520557} \\
 \midrule
 $f_{1}$ & 9.240112(669) & 0.23758(62) & 5.8628(238) & 382 & 0.00087  \\
% $f_{2}$ & 0.041719(739) & 0.46608(149) & 3.3411(263) & 314 & 0.00085  \\
 $f_{2}$ & 0.367199(738) & 0.19113(61) & 1.6745(263) & 314 & 0.00084  \\
 $f_{3}$ & 0.992114(1101) & 0.14742(104) & 0.0614(392) & 141 & 0.00083  \\
 $f_{4}$ & 2.965851(1358) & 0.11024(119) & 2.2809(484) & 93 & 0.00080  \\
% $f_{6}$ & 86.507851(5110) & 0.02694(411) & 0.0369(1821) & 7 & 0.00075  \\
% $f_{7}$ & 321.709440(5790) & 0.02368(464) & 1.1401(2063) & 5 & 0.00075  \\
\midrule
 \multicolumn{5}{l}{TIC 93997745} \\
 \midrule
 $f_{1}$ & 0.893092(317) & 1.37146(35) & 5.7259(74) & 3918 & 0.00122  \\
 $f_{2}$ & 12.466929(681) & 0.37048(44) & 1.0048(160) & 849 & 0.00077  \\
 $f_{3}$ & 0.860712(849) & 0.31497(58) & 1.2051(199) & 547 & 0.00072  \\
 $f_{4}$ & 12.776816(866) & 0.27466(52) & 0.6852(203) & 526 & 0.00069  \\
 $f_{5}$ & 0.167591(1181) & 0.37780(134) & 1.4704(277) & 283 & 0.00063  \\
\midrule
 \multicolumn{5}{l}{TIC 339570153} \\
 \midrule
 $f_{1}$ & 8.334652(32) & 1.79744(110) & 3.0236(115) & 1639 & 0.00284  \\
 $f_{2}$ & 9.849075(36) & 1.58846(117) & 5.0767(127) & 1355 & 0.00250  \\
 $f_{3}$ & 13.152522(45) & 1.11946(130) & 5.7441(159) & 863 & 0.00225  \\
 $f_{4}$ & 12.307331(49) & 0.89725(126) & 3.7999(175) & 711 & 0.00211  \\
 $f_{5}$ & 1.602285(49) & 1.00083(137) & 3.4750(172) & 732 & 0.00200  \\
 $f_{6}$ & 0.158870(55) & 1.25039(218) & 5.7760(194) & 575 & 0.00189  \\
% $f_{7}$ & 24.620136(65) & 0.60297(146) & 1.1020(229) & 414 & 0.00182  \\
\midrule
 \multicolumn{5}{l}{TIC 264540478} \\
 \midrule
% $f_{1}$ & 0.074269(1927) & 1.53971(326) & 3.9231(214) & 473 & 0.00159  \\
 $f_{1}$ & 0.785078(2079) & 1.03330(254) & 4.1115(231) & 406 & 0.00146  \\
 $f_{2}$ & 11.181360(1910) & 0.66533(138) & 1.0108(212) & 481 & 0.00125  \\
 $f_{3}$ & 8.928850(2377) & 0.53031(171) & 0.3638(264) & 311 & 0.00114  \\
 $f_{4}$ & 1.190526(2496) & 0.55627(197) & 3.6939(278) & 282 & 0.00108  \\
\midrule
 \multicolumn{5}{l}{TIC 379012185} \\
 \midrule
 $f_{1}$ & 1.571877(1895) & 2.68020(697) & 5.9813(238) & 384 & 0.00604  \\
 $f_{2}$ & 1.446061(1735) & 2.81841(615) & 0.9768(218) & 458 & 0.00575  \\
 $f_{3}$ & 0.637746(1683) & 2.64853(544) & 5.9925(211) & 487 & 0.00511  \\
 $f_{4}$ & 6.141675(1855) & 1.93608(483) & 2.1626(233) & 401 & 0.00447  \\
 $f_{5}$ & 1.359406(1834) & 2.19885(536) & 4.7540(230) & 410 & 0.00425
\end{longtable}

%%%%%%%%%%%%%%%%%%%%%%%%%%%%%%%%%%%%%%%%%%%%%%%%%%%%%%
% Don't change these lines
\bsp	% typesetting comment
\label{lastpage}
\end{document}